\documentclass[aps,twocolumn,showpacs,superscriptaddress]{revtex4-1}
\usepackage{amsmath,amssymb}
\usepackage{graphics,graphicx}
\usepackage{dcolumn,bm}
\usepackage{psfrag}

\newcommand{\cu}
{\affiliation{Department of Physics, University of Calcutta,
92 Acharya Prafulla Chandra Road, Kolkata 700009, India.}}

\begin{document}

\title{Continuous utility factor in segregation models}

\author{Parna Roy}
\cu
\author{Parongama Sen}
\cu

\begin{abstract}

We consider the constrained Schelling model of social segregation in which the utility factor of agents strictly increases 
and non-local jumps of the agents are allowed. In the present study, the utility factor $u$ is defined in a way 
such that it can take continuous values and depends on the tolerance threshold as well as the fraction of unlike neighbours. 
Two models are proposed: in model A the jump probability is determined by the sign of $u$ only which makes it equivalent to the discrete model. 
In model B the actual values of $u$ are considered. Model A and model B are shown to differ drastically as far as 
segregation behaviour and phase transitions are concerned. In model A, although segregation can be achieved, the cluster sizes 
are rather small. Also, a frozen state is obtained in which steady states comprise of many unsatisfied agents. 
In model B, segregated states with much larger cluster sizes are obtained. 
The correlation function is calculated to show quantitatively that larger clusters occur in model B.
Moreover for model B, no frozen states exist
even for very low dilution and small tolerance parameter. 
This is in contrast to the unconstrained discrete model considered earlier where agents can move even when utility remains same. 
In addition, we also consider a few other 
dynamical aspects which have not been studied in segregation models earlier. 

\end{abstract}


\maketitle

\section{Introduction}
Several dynamical phenomena are associated with the mobility of human beings. Pedestrian's movement patterns \cite{helbing_review} and 
associated  behaviour,  synchronized behaviour 
such as flocking or herding \cite{vicsek} are some well known examples.
Mobility of human beings is often motivated by some definite purpose and emergent phenomena 
such as segregation and migration may result \cite{pb}. 
Dynamical phenomena associated with movement of human beings and animals have attracted the attention of physicists as
many of these can be mapped to dynamical physical systems. For example, pedestrians' movements 
can be analysed using a social force model \cite{helbing_review} and  flocking phenomena can be studied in terms of the classical XY model \cite{vicsek}.
   
The segregation phenomena is fundamental in nature and seen to be present in the human society 
in different forms. 
In a heterogeneous society, consisting of different racial or 
cultural groups, it is often found that people prefer to live in the neighborhood of people belonging to 
the same group \cite{smith} which results in a segregation. Segregation may occur due to different factors such as 
age, income, language, religion, color etc. To achieve segregation in space, one must consider the movements 
of the agents following certain rules. A simple model of social segregation was proposed by Schelling \cite{sch1,sch2,stauffer2} 
long before the issue of residential segregation received serious attention \cite{clark}. This 
model illustrates how an individual's choice of living with the 
neighbor of the same kind can lead collectively to segregation. 
The model received attention from physicists as it was realised that
its continuous analog, in space and time,  
corresponds to a  dynamical model  of  solid or liquid flow  \cite{kirman}.
On the other hand, it has similarity with magnetic models with conservative dynamics \cite{Meyer,Grauwin} 
and also the dynamical phenomena of phase ordering \cite{bray}.
Later, several variants of the Schelling model were studied. In \cite{rogers} a unified mathematical framework was developed 
for a broad class of models of the Schelling type. In most of the Schelling class models, agents occupy sites of a two dimensional grid. 
In \cite{mckane} it was considered that the agents occupy districts (or patches) rather than 
sites on a two dimensional grid, which improves social realism of the model.

In the original Schelling model, 
a  utility factor is defined based on the fraction 
of opposing neighbours; if the latter is greater (less) than $1/2$, the utility factor is  defined to be zero (one).
An unsatisfied agent with zero utility tends to move to a vacant site in the neighbourhood 
only if that makes the utility factor larger (constrained dynamics). 
However, it was shown later that this leads to segregated clusters which are rather 
small in size.
It was found that  the system can  reach a segregated state where large clusters are formed
provided
agents
are allowed to move even when the utility remains same \cite{kirman,stauffer,asta} - this is called the unconstrained case. 
The configurations generated in this way are very similar to those occurring in the phase separation dynamics.

Whether the dynamics is constrained or unconstrained
is 
a major concern in the Schelling model.
In  \cite{asta} both the constrained and unconstrained cases were considered in one dimension and  it was shown that the constrained model
shows non-trivial static properties characterised by the presence of a symmetry breaking phase transition. On the other hand, in
the unconstrained model the dynamics exhibit coarsening as in Ising-like models with non-conserved order parameter.
Also, it was shown in \cite{kirman} that the constrained case corresponds to 
``solid-like" flow while the unconstrained case mimics the flow of liquids. 

In the present paper, we consider continuous values of the utility factor (details in the next section) and 
 show that even subject to  the constrained rule (that unsatisfied agents can move to a different location only if their utility increases),
 it is possible to obtain good quality segregation, i.e segregated clusters are larger in size. 
 This is comparable to the unconstrained case with discrete utility factors.

 The case of continuous utility was considered earlier in \cite{benenson,hatna} with added degrees of freedom 
 and the effect of continuous utility alone was not obvious. 
In \cite{pancs} utility was regarded as a general function of $f$. But only the case where $F = 1/2$ was 
taken and a special form (spiked) of the utility function was used (utility equal to 1 for $f=1/2$ and zero otherwise). 
In  \cite{kirman},
where the mapping of the Schelling model to a physical model  in continuum
had been made, 
 the utility factor was also taken as a continuous function. 
However, results were obtained only for the case where it has
a step function like behaviour.

The reason for considering utility as a continuous valued variable is twofold.
First, it gives an idea of the degree of mismatch between the agent and their neigbhours. Secondly,  it provides a greater mobility to the 
unsatisfied agents
as they can now move to a different location even when utility does not become
positive necessarily but definitely improves. This is close to 
reality as it is not possible to attain an ideal state always; especially in a single step.
So in an indirect 
manner, the movements of agents become less constrained in the continuous case. 

Apart from studying the segregation phenomena (in terms of satisfied agents) and related phase transitions, certain dynamical 
features are also studied for the two models. 
%Moreover, we show that phase transitions occurring in the constrained case with discrete and continuous utility
%factors are quite different in nature.

In section II we have defined the models and the quantities calculated and section III gives the details of the simulation. 
In section IV results 
are discussed, section V contains study of the correlation function and section VI contains summary and discussions.

\section{Models and quantities calculated}

First, we describe the Schelling model with discrete utility factor where 
the agents belonging to two different groups are located on the sites 
of a chessboard. Some sites are left vacant, the fraction of vacant sites is denoted by the dilution parameter $p$. 
The neighborhood of an agent comprises
eight nearest sites (Moore neighborhood). The agents are able to relocate according to the fraction of neighboring agents belonging to 
their own group. 
An agent located at the center of a neighborhood where the fraction of neighbors of opposite group (denoted by $f$) 
is greater than a predefined tolerance threshold $F$, will try to relocate to a neighborhood for which  $f$ 
 is less than or equal to  $F$  (in the original model, $F = \frac{1}{2}$). 
The agents for which $f\le F$, utility factor $u$ is defined to be equal to $1$ and the agent is 
 said to be satisfied. Otherwise the agent is unsatisfied and utility factor is $0$.

In this paper, we have introduced the utility factor as 
\begin{equation}
 u=F-f,
 \label{eq1}
\end{equation}
 which can essentially take continuous values. $f$ can take  values between 
$0$ and $1$ %(however, it is not possible to span the entire space as $f$ can take a finite number of values) 
 and $0 \le F \le 1$. In principle, $f$ takes a finite number of values, e.g. in 
 a Moore neighborhood $f$ can have $23$ different values. These values are evenly spread 
 in the interval [0,1]. Compared to the original model where $u$ is binary, $u$ can assume a much larger number of values, 
 evenly spread in the interval [$F-1$,$F$] and hence we regard it to be continuously varying.
 We allow here realistic constrained nonlocal jumps, i.e.
 movement to vacant sites where utility can be increased only for the unsatisfied agents who have $u<0$. Satisfied agents do not need to move. 
 We consider here two models: model A and model B.
  In model A, an unsatisfied agent can move to any randomly chosen vacant site where $u \ge 0$. In model B 
  an unsatisfied agent can move to any randomly chosen empty site provided $u$ has a larger value compared to the original one. 
  Effectively model A is equivalent to the original model where $u$ is discrete as only 
  the sign of $u$ matters during a movement. 
However, it is useful to study model A to directly compare the results of model B with continuous utility factors.
We check how the segregation 
  is affected by varying the dilution parameter $p$ and tolerance threshold $F$ in these two models. 
  A similar study was done in \cite{gauvin}, with $u$ taking discrete values only.

We take a  $L \times L$ lattice 
with  a fraction $p$ of sites  empty at random. 
Agents belonging to two different groups (of equal size) occupy the rest of the  sites randomly in the beginning. 
If $N_{A}$ and $N_{B}$ denote the number of agents belonging to two different groups, then $N_{A}+N_{B}=N$, $N_{A}=N/2$, $N_{B}=N/2$ 
and $N=(1-p)L^{2}$.
The fraction of neighbors belonging to the opposite group 
 $f$ is calculated by  
  dividing the number of opposing neighbours  by the total number of occupied neighboring sites. 
  For example if $6$ neighboring sites are occupied and $2$ of them belong 
  to the opposite group then $f=\frac{1}{3}$. The total number of neighbors is a variable 
 and can vary from $0$ to $8$. If the total number of neighbors of any agent is $0$ then the agent 
 is taken as satisfied, i.e., they have a positive utility.

 Segregation models are known to exhibit phase transitions. 
One can vary factors like the threshold value $F$, the dilution factor $p$, size of the two groups  etc. 
to investigate  the presence of a  critical value below/above  which 
segregation can occur. 
As already mentioned, in \cite{asta}, a phase transition in the constrained case was observed as  the dilution factor was increased.  
In the unconstrained model considered in \cite{gauvin}, where 
satisfied agents could also move, it was found that by varying  $p$ and $F$, 
a phase diagram can be obtained. There are two transitions; for very small $F$ the states are frozen, increasing $F$ 
one can get a segregated state while above a critical $F$, a mixed state exists.

In the present model, we calculate two important quantities. The first is $\phi$, the fraction of agents with negative utility. 
If $m(u)$ denotes the fraction of population with utility $u$,
\begin{equation}
 \phi=\sum_{u<0} m(u).
 \label{eq2}
\end{equation}
$\phi$ is zero for $F=1$ trivially. 
The other quantity is  the average utility  
defined as $\langle u\rangle =\langle F-f \rangle$.

The average fraction of opposing neighbours $f_{avg}$ helps in understanding 
qualitatively whether segregation is reached. One can define 
\begin{equation}
 s=1-f_{avg} 
\end{equation}
to be the segregation factor; the larger is $s$, the better is the degree of segregation. 
Three types of states may occur; the frozen state where $\phi$ has large non-zero value and $s$ is very close to $0.5$ 
(which is the value expected for a completely homogeneous distribution of agents which is initially chosen). 
This situation is analogous to the jamming transition \cite{mckane} in segregation model in
which large numbers of agents remain stuck in unfavorable states. As $F$ is made larger, 
$s$ may increase. If $s$ is very close to $1$ and $\phi$ is negligible, one may call that a segregated state. For even larger $F$, 
$s$ will decrease from its maximum value while $\phi=0$; eventually for $F=1$, $s=0.5$ and $\phi=0$, which is of course a mixed state. 
Our goal is to study whether there are sharp or smooth transitions from one state to another as $F$ is increased.

We have also calculated some dynamical
 quantities  related to the motion of the agents. 
   The first is the  persistence probability $p_{move}$, which 
is defined as the fraction of agents who have not moved till time $t$.
Another quantity calculated is the  mobility factor defined as the  
average distance travelled by the agents, denoted by $d$. This is the average Euclidean distance between the initial and final positions (position 
after reaching steady state) of the agents. 
 These two factors have not been studied in the context of segregation models earlier, 
 to the best of our knowledge.
 
\section{Details of simulations}
In the initial configuration the $N$ agents belonging to the two different groups are homogeneously distributed among the $L^{2}(1-p)$ sites. 
A site is chosen at random at every time step. 
If the chosen site is occupied then it is tested whether the agents are satisfied or not. If the 
agents are satisfied they stay at their present site, otherwise they select one unoccupied site at random from all sites that are unoccupied at that moment. 
We assume that the list of unoccupied locations is available to the agents. The unsatisfied agents move to the 
randomly chosen empty site provided they become satisfied there (model A) or the utility factor increases (model B). Otherwise they stay at 
their present site. Only one such attempt is allowed. Dynamics stop when the system reaches an absorbing state; i.e there is no change in the 
dynamical quantities measured. An absorbing state is always possible here. We have used $p$ values from $0.02$ to $0.6$ and $0<F<1$. For each set of parameters $2500$ initial configurations 
are used over which the relevant quantities are averaged.                                                                                                                         We have used different $L$ values; results for $L=20$, $30$ and $50$ are presented.

\section{Results}
\subsection{Steady state behaviour and phase diagram}
%\subsection*{Model A}

{\it {Model A}}:
 We have studied the time dependence of various quantities defined in the last section and observed 
that they all reach a steady state value in time. 

\begin{figure}[!htbp]
 \includegraphics[width=7.5cm,height=5.0cm,angle=0]{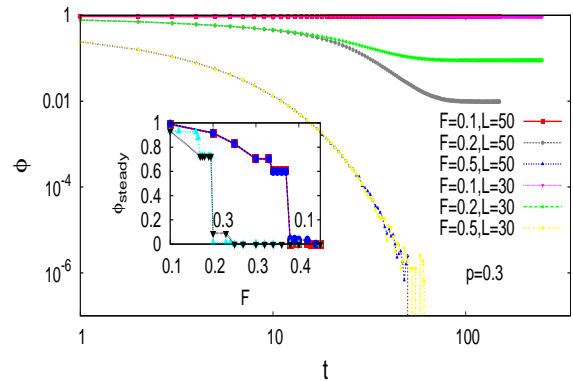}
  \caption{(Color online) Model A: $\phi$ (fraction of agents with utility factor $u<0$) for $p=0.3$ for $L=30$ and $L=50$ 
  for several values of $F$ against time.
  Inset shows $\phi$ at steady state for $p=0.3$ and $p=0.1$ against $F$. Results have negligible system size dependence.}
  \label{size1}
\end{figure}
 \begin{figure}[!htbp]
 \includegraphics[width=7.5cm,height =5.0cm,angle=0]{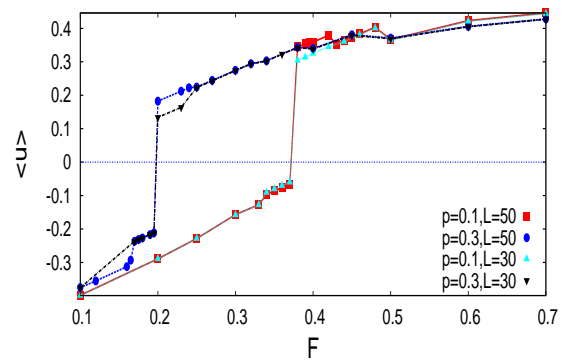}
  \caption{(Color online) Model A: Average value of $u$ at steady state for $p=0.3$ and $p=0.1$ for $L=30$ and $L=50$ 
  against $F$. Results have negligible system size dependence.}
 \label{jump3}  
  \end{figure} 
  
  \begin{figure}[!htbp]
 \includegraphics[width=7.5cm,height=5.0cm,angle=0]{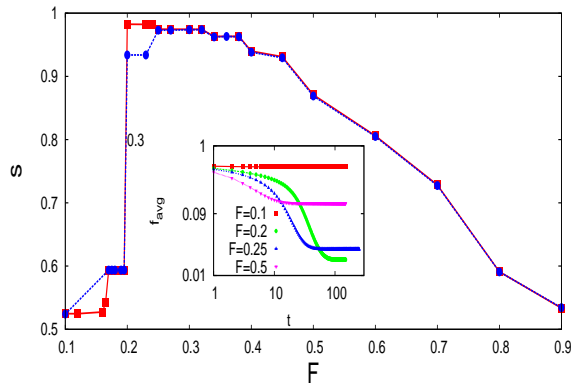}
 \caption{(Color online) Model A: Segregation factor at steady state for
 $p=0.3$ for $L=30$ and $L=50$ against $F$. Inset shows time variation of $f_{avg}$ (average fraction of opposite neighbor)
 for $p=0.3$ for $L=50$. }
 \label{dist}
\end{figure} 
 
We first discuss the behaviour of $\phi$ (eq. \ref{eq2}) as a function of time. 
We find that in general for small $F$, $\phi$ reaches a nonzero saturation value but for larger values of $F$, the saturation value 
of $\phi$ becomes zero (Fig. \ref{size1} shown for $p=0.3$). 
Plotting the saturation values (inset of Fig. \ref{size1}), we find that there is a critical value of $F = F_{c}$
where $\phi$ decreases to a negligible value ($\approx 0$) discontinuously.
Hence this critical value separates two regions of $\phi=0$ 
and $\phi \neq 0$. $F_c$ depends on $p$. 
%It may be noted here that the data do not show any system size dependence.

Next we discuss the behaviour of $\langle u \rangle$ (average value of $u$) in the steady state.
As can be seen from Fig. \ref{jump3}, this goes from a negative 
value to a positive value as $F$ is made larger and crosses zero at a value of 
$F \simeq F_{c}$. In analogy with magnetic or liquid-gas 
phase transition, one can interpret the $\langle u \rangle =0 $ point as 
a coexistence point. Thus $F_{c}$ acts as a field which separates 
the two regions with $\langle u \rangle < 0$ and $\langle u \rangle > 0$  much like an 
external field in magnetic systems below the critical point. 

For complete characterisation of the system the variation of the saturation value of $s$, the segregation factor 
as a function of $F$ is studied.  We find that $s$ sharply increases 
 from $s\approx0.5$ to a large value ($\approx1$) at a value of $F\simeq F_{c}$. 
 Hence we find a region where $\phi>0$ and $s\approx0.5$ which we identify as the frozen state. As $\phi$ drops sharply 
 to zero at $F_{c}$ and $s$ also shows a sharp increase at the same point, one can 
 conclude that a transition between a frozen state and a segregated state occurs at $F_{c}$. 
 Since there is hardly any finite size dependence in the data and all the quantities show 
 sharp changes at $F_c$ we conclude that the transition is first order in nature.

\begin{figure}[!htbp]
  \includegraphics[width=8.0cm,height=5.2cm,angle=0]{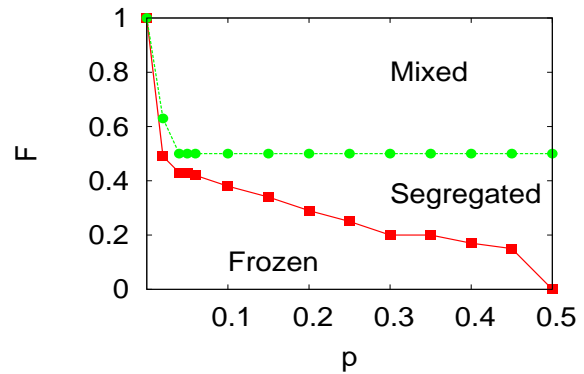}
  \caption{(Color online) Model A: Phase diagram of model A. The lower line represents transition between frozen and mixed state. 
  The upper line seperates segregated state and mixed state. Continuous lines guide to the eye.}
  \label{phase1}
 \end{figure}

Apparently, no sharp transition occurs between the segregated state and mixed state as $s$ decreases 
for larger values of $F$ approaching $\simeq0.5$ as $F\to 1$ monotonically. However we note that $s$ remains almost constant 
for a range of value of $F$ before decreasing. This suggests there may be another transition occurring at $F>F_c$ between 
a segregated state and a mixed state. As it is difficult to identify 
the second transition point, we define a ``transition point'' where $s$ becomes less than $0.9$, 
i.e we define the state as segregated when $s \ge 0.9$.  
Based on these findings, we show the transition points between the three different phases in the $p-F$ plane for Model A in Fig. \ref{phase1}.                                                                                                                                                                                                                                                                                                                                                                                                                                                                                                                                                              

\begin{figure}[!htbp]
  \includegraphics[width=8.0cm,height=5.2cm,angle=0]{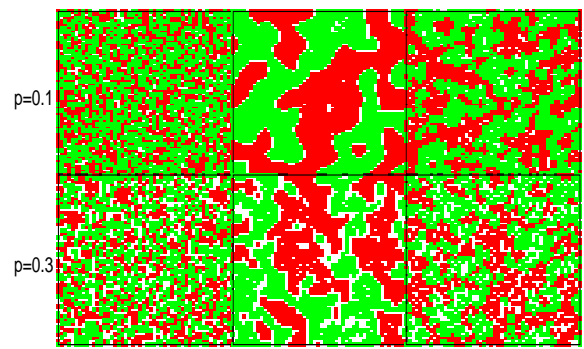}
  \caption{(Color online) Model A: Snapshots of the steady states at $p=0.1$ (upper panel) and $p=0.3$ (lower panel). For $p=0.1$, snapshots are shown 
  for $F=0.1$, $F=0.4$ and $F=0.7$ corresponding to frozen state, segregated state and mixed state respectively. For $p=0.3$, snapshots are shown 
  for $F=0.1$, $F=0.3$ and $F=0.7$ corresponding to frozen, segregated and mixed state respectively.
  The white regions correspond to blank spaces and the two different shaded regions correspond to two different groups which occur with same probability.}
  \label{snap1}
 \end{figure}

 We next present some typical snapshots 
 in Fig. \ref{snap1} 
to show the effect of increasing $F$   
and the quality of segregation. 
The snapshots show the  steady
states  for $p=0.1$ and $p=0.3$.
For a small value of tolerance 
threshold $F$, the probability that an agent is unsatisfied is quite large. 
However, the probability that a favourable site be found by  relocation is also small at small $F$ (unless $p$ is sufficiently
large so that the agent has more choice of making movements). For a more detailed discussion see section VI. 
 Thus the initial grey picture where the agents of two groups are mixed up remains almost unchanged i.e., frozen in time.
 
As $F$ is made 
larger, fewer agents will be unsatisfied, however, the probability to find an empty site which will satisfy the agent is larger and the system will undergo a time evolution 
and the steady state picture will show finite sized domains of agents of the same group making $s>0.5$. 
For smaller value of $p$, the effect will show up for a larger value of $F$. 
If $F$ is made even larger, 
the need for relocation becomes less since now almost all the agents are satisfied. Although some agents will move, 
the cluster sizes will be much less. This is apparent in the figures for $F=0.7$ (Fig. \ref{snap1}). 

We note that the snapshots are very similar to those  of the constrained 
Schelling model (see for example \cite{kirman}) which is consistent with the fact that model A effectively uses binary  utility factors.  
The cluster sizes, even in the so called segregated state, are small with the interface showing a lot of roughness.

{\it {Model B}}:
%\subsection*{Model B}
  In model B also we have studied the time dependence of the relevant quantities and observed that they
   reach a steady state value in time. 

\begin{figure}
 \includegraphics[width=7.5cm,height=5.0cm,angle=0]{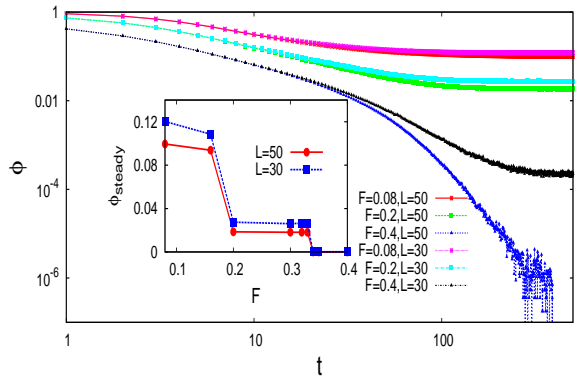}
  \caption{(Color online) Model B: $\phi$  for $p=0.1$ for $L=30$ and $L=50$ against time for several values of $F$.
   Results show appreciable system size dependence for smaller $F$. Inset shows $\phi$ at steady state for $p=0.1$.}
  \label{njump1} 
 \end{figure}

  The most striking result in model B is that a nonzero value of $\phi$ is  obtained only for 
  rather small values of $p$ ($\le0.2$) for 
  any value of $F>0$.   
In  Fig.  \ref{njump1} we have plotted $\phi$ as a function of time for $p=0.1$. 
   As in model A, here also $\phi$ reaches a nonzero saturation value for smaller $F$ and goes to zero for larger $F$ (shown in the inset of Fig. \ref{njump1}). 
   Although the plot of saturation values 
   of $\phi$ with $F$ lacks smoothness, there are two clear indications, first, the saturation values of $\phi$ are rather small even for small values of $F$. 
   Secondly, there is appreciable system size dependence. In fact we find that $\phi$ decreases with system size 
   and as the $\phi$ values are already $\lesssim 0.1$ for $L=50$, one can conjecture that $\phi$ would 
   vanish for any $F$ in the thermodynamic limit (see Fig. \ref{njump1} inset).

\begin{figure}[!htbp]
 \includegraphics[width=7.5cm,height=5.0cm,angle=0]{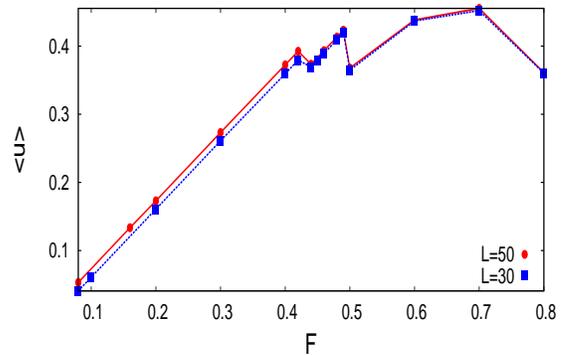}
  \caption{(Color online) Model B: Average utility factor $u$ at steady state for $p=0.05$ for $L=30$ and $L=50$. Results 
  show system size dependence for small $F$ values.}
 \label{njump3}  
  \end{figure}

   The behaviour of $\langle u \rangle$ in the steady state (Fig. \ref{njump3}) is also completely  different from model A and 
   supports the conjecture that $\phi$ vanishes for all $F$ and $p$.
   In model A,  the steady state value of $\langle u \rangle$ goes from a negative to a positive value sharply at a particular 
   value of $F$ which is dependent on $p$. But in model B, the steady state value of $\langle u \rangle$ is 
   always positive no matter how small $F$ is ($F>0$) and it shows finite size dependence. $\langle u \rangle$ is larger 
   for a larger system size; although above $F\approx0.5$, system size dependence is negligible.
Hence the value of $F_{c}$ (where $\langle u \rangle$ turns positive in model A) is simply equal to zero for model B.
Physically this implies that there is no frozen state in model B (see section VI for an argument why this happens). 

  
\begin{figure}[!htbp]
%\hspace{-1.5cm}
 \includegraphics[width=7.5cm,height=5.0cm,angle=0]{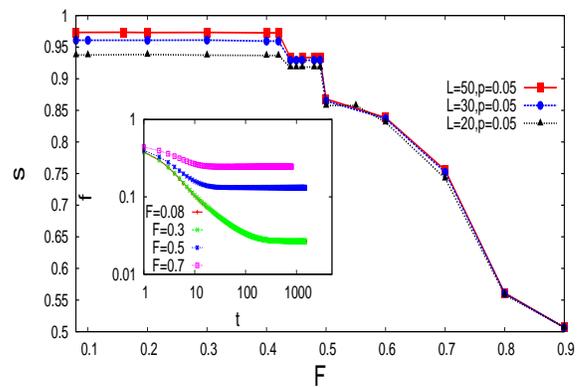}
  \caption{(Color online) Model B: Segregation factor $s$ at steady state for $p=0.05$ for $L=20$, $L=30$ and $L=50$ against $F$. 
  Results show considerable system size dependence and $s$ increases with system size.
   Inset shows time variation of $s$ for $p=0.05$ for $L=50$ for $F=0.08$, $0.3$, $0.5$, $0.7$ respectively from bottom to top respectively.}
  \label{nfig3}
\end{figure}

In Fig. \ref{nfig3} we have shown the behaviour of segregation factor $s$. From this 
figure we can see that the steady state value of $s$ is much larger than $0.5$ even for small $F$ and remains almost constant over 
a considerable interval of $F$ before decreasing.
The decrease for large $F$ values is due to the same reason as in model A; 
the agents are satisfied when $F$ is large even with a considerably large value of $f$. $s$ also shows appreciable system size 
dependence, it increases with system size and approaches $1$ in the thermodynamic limit for smaller values of $F$.

\begin{figure}[!htbp]
 \includegraphics[width=8.0cm,height=5.2cm,angle=0]{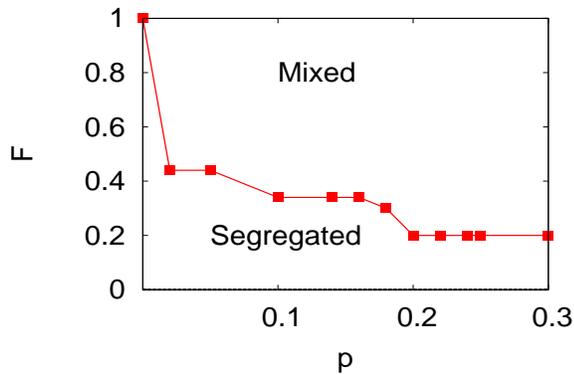}
  \caption{(Color online) Model B: Phase diagram of model B with only one transition line for nonzero $F$. 
  It represents transition between segregated and mixed state. $F=0$ line represents transition between frozen and 
  segregated state. Continuous line guides to the eye.}
  \label{phase2}
\end{figure}

We show in Fig. \ref{phase2} the transition points between the segregated and mixed phase for model B. Since in model B there is no frozen state, 
there is only one transition line for non-zero $F$. As in model A, there is no sharp transition between the 
segregated and mixed state and the boundary between the 
segregated state and mixed state is obtained using the same criteria as in model A. 

\begin{figure}[!htbp]
 \includegraphics[width=8.0cm,height=5.2cm,angle=0]{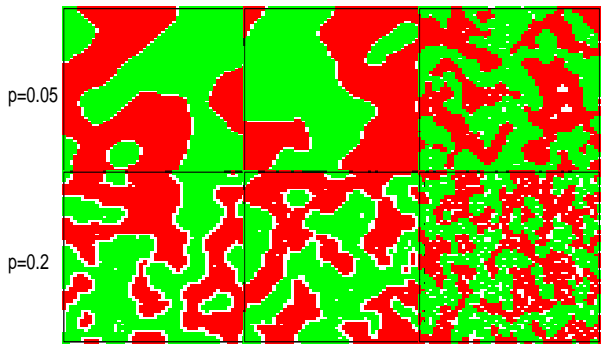}
  \caption{(Color online) Model B: Snapshots of the steady states at $p=0.05$ (upper panel) and $p=0.2$ (lower panel)
  for tolerance $0.08$, $0.4$ and $0.6$ respectively. $F=0.08$ and $F=0.4$ corresponding to segregated state and $F=0.6$ corresponding to 
  mixed state. The white regions correspond to blank spaces and the two different shaded regions correspond to two 
  different groups which occur with same probability.}
  \label{nsnap}
\end{figure}

Typical snapshots of the steady states for model B are shown in Fig. \ref{nsnap}.
The steady state snapshots are almost identical in nature to those occurring for 
the unconstrained case with discrete utility factors \cite{kirman} 
and hence we find that even with the constrained case, it is possible
to obtain segregated states with large cluster formation provided the utility factors are continuous. 
The picture for small $F$ is very similar to that occurring in phase seperation dynamics.
We note that the cluster sizes are much larger compared to those in model A especially for small $F$ where $s$ 
is close to $1$. Also the interfaces are much more smooth
and the overall scenario is similar to liquid flow dynamics. Even for larger values of $F$ where 
the mixed state occurs, the cluster sizes are apparently larger compared to model A.
 
\subsection{Results related to mobility}
%\subsection*{Model A}

We have evaluated persistence probability and average distance travelled in both the models. 
%In particular we are interested to see the behaviour of these quantities close to the phase boundary.

\begin{figure}[!htbp]
 \includegraphics[width=7.5cm,height=5.0cm,angle=0]{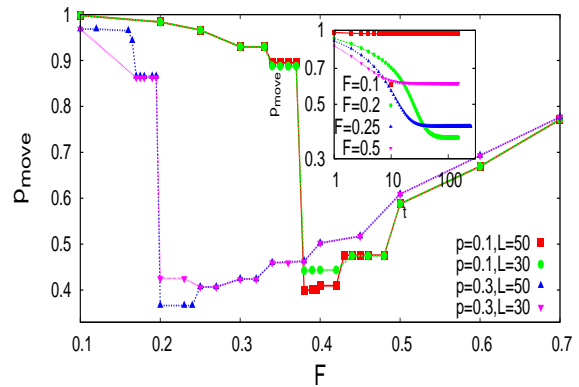}
  \caption{(Color online) Model A: plot of persistence probability $p_{move}$ corresponding to movement at steady state for $p=0.3$ and $p=0.1$.
  Inset shows time dependence of persistence for $p=0.3$.}
  \label{jump2}
 \end{figure}

{\it Model A}:
We have plotted the persistence probability $p_{move}$ corresponding to movement of the agents  as a function of time in the inset of Fig. \ref{jump2}. The 
main plot shows the steady state values which drop to a considerably smaller value close to $F_c$. There is also a non-monotonic 
behaviour of the steady state values as a function of $F$. 
The plots support the picture that for small $p$, the agents show very little movement leading to frozen states. Once again 
system size dependence is negligible.

\begin{figure}[!htbp]
 \includegraphics[width=8.5cm,height=4.8cm,angle=0]{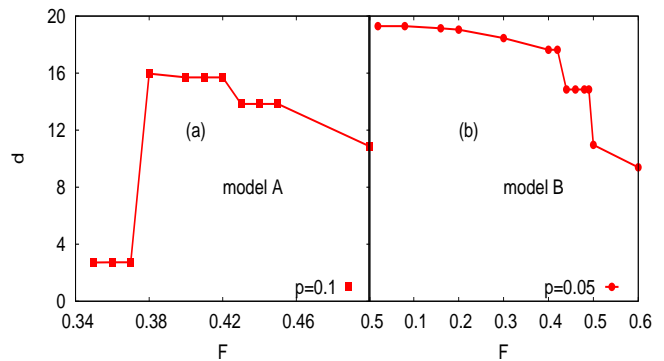}
 \caption{(Color online) Plot of average distance moved by each agents at different tolerance for 
 $p=0.1$ for model A (a) and $p=0.05$ for model B (b).}
 \label{dist2}
\end{figure}

The plot of average distance travelled by the agents $d$ against $F$ are shown in Fig. \ref{dist2}a. From this figure one can see that the average 
distance moved by the  agents suddenly increases close to  $F=F_{c}$ and 
then slowly decays with $F$.

\begin{figure}[!htbp]
 \includegraphics[width=7.5cm,height=5.0cm,angle=0]{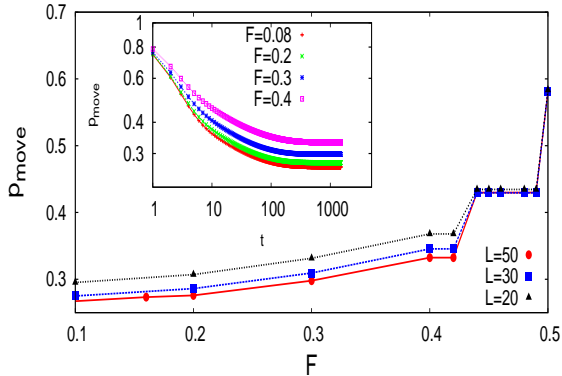}
  \caption{(Color online) Model B: plot of persistence probability $p_{move}$ corresponding to movement at steady state for $p=0.05$ for $L=20$, $L=30$ and $L=50$. 
  Results are system size dependent. Inset shows time variation of persistence for $p=0.05$ for $L=50$ for $F=0.08$, $0.2$, $0.3$, $0.4$ respectively from bottom to top.}
  \label{njump2}
 \end{figure}

{\it Model B}: For model B  
  the  persistence probability corresponding to movement of agents with time is plotted in   Fig. \ref{njump2}.  
 We have plotted the steady state value of $p_{move}$  for a rather 
 small value of $p=0.05$ and have found that there is  a monotonic increase as a function of $F$.
Also, the steady state probability is fairly low for $F \lesssim 0.4$ signifying the agents have high mobility.  
Considerable system size effects are present; for larger systems, the persistence probability decreases for $F \lesssim 0.4$.
 
 In Fig. \ref {dist2}b the average distance travelled  by 
the  agents against $F$ is plotted.  There is an  overall tendency of a  slow decay as $F$ is increased. 

The behaviour of $p_{move}$ and $d$ can be easily explained as $F$ is made larger; the need to move decreases as the agents become more tolerant and hence
there is less movement making $p_{move}$  large and $d$ small. This is true in both the models.

%For small $F$ it has considerably larger values 
% which shows slow decrease for $F$ values which is not too large and then considerably 
% decreases as $F$ is increased beyond $F_{c1}$. This fact is consistent with the behaviour of $f$ (Fig. \ref{nfig3}). 

%We thus find that the effect of a phase transition at $F_c$ is strongly felt in
%case of Model A where $p_{move}$  ($d$) shows a sudden decrease (increase). 
%No such effect is noted in Model B close to $F_{c1}$.

\section{Correlation}
\begin{figure}[!htbp]
 \includegraphics[width=7.5cm,height=5.0cm,angle=0]{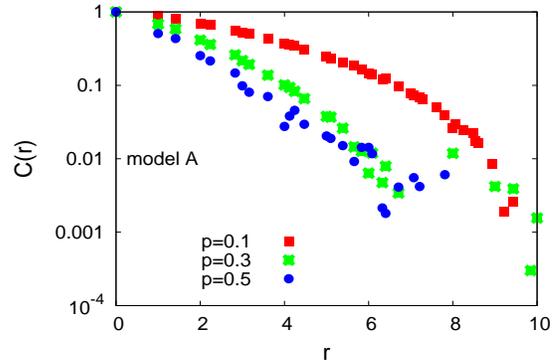}
  \caption{(Color online) Model A: Decay of correlation with distance $r$ for $p=0.1$, $0.3$ and $0.5$ for $F=0.4$.}
  \label{corrA}
 \end{figure}

\begin{figure}[!htbp]
 \includegraphics[width=7.5cm,height=5.0cm,angle=0]{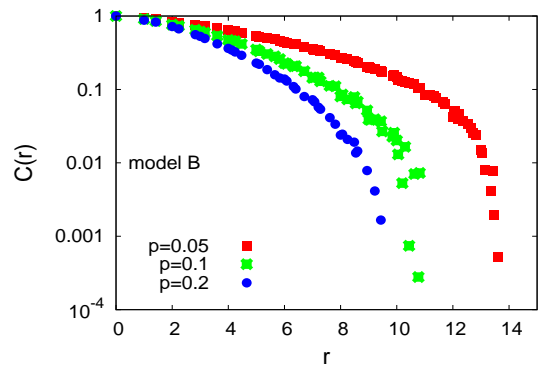}
  \caption{(Color online) Model B: Decay of correlation with distance $r$ for $p=0.05$, $0.1$ and $0.2$ for $F=0.05$.}
  \label{corrB}
 \end{figure}

\begin{figure}[!htbp]
 \includegraphics[width=7.5cm,height=5.0cm,angle=0]{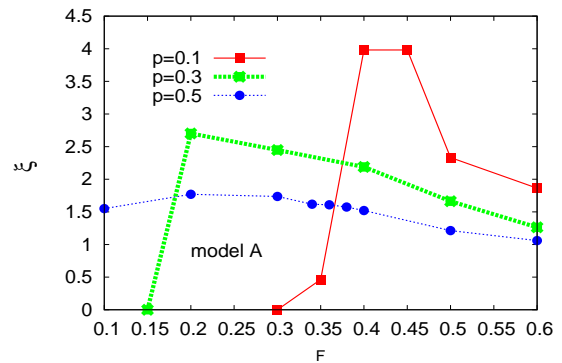}
  \caption{(Color online) Model A: Plot of effective correlation length $\xi$ with $F$ for $p=0.1$, $0.3$ and $0.5$.}
  \label{fitA}
 \end{figure}

\begin{figure}[!htbp]
 \includegraphics[width=7.5cm,height=5.0cm,angle=0]{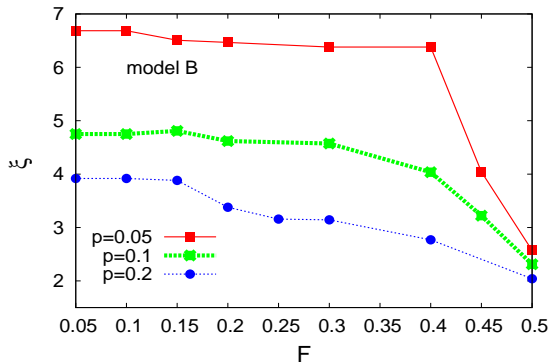}
  \caption{(Color online) Model A: Decay of correlation with distance $r$ for $p=0.05$, $0.1$ and $0.1$.}
  \label{fitB}
 \end{figure}

 We have calculated the correlation as a function of distance in the steady state to have an idea of the cluster size, the latter is a measure 
 of the quality of segregation. If the $i$th site is occupied 
 we have assigned values $S_i=\pm1$ corresponding to the two different groups; if it is empty $S_i=0$. 
 Correlation between two sites 
 at a distance $r$ is defined as $C(r)=\langle S_iS_j\rangle$ where $r$ is the Euclidean distances between the $i$th and $j$th sites. Figures \ref{corrA} and \ref{corrB} show the decay of correlation $C(r)$ 
 with distance $r$ for model A and model B respectively. The data is fit with the form $C(r)\sim \exp(-(\frac{r}{\xi})^a)$ and  
 the effective length scale $\xi$ upto which the sites are correlated for different $F$ for both the models is extracted. Values 
 of $a$ depend on the parameters. Figures \ref{fitA} and \ref{fitB} 
 show the plots of $\xi$ as a function of $F$ for model A and model B respectively. From these figures it is evident that 
 the cluster sizes are larger in model B compared to model A, i.e, it is possible to obtain a good quality segregation even 
 in the constrained model with continuous utility.

\section{Summary and discussions}

In this paper we have defined continuous utility factors in the constrained 
Schelling model with non-local jumps. 
We have introduced utility factor as $u=F-f$ where $F$ is the tolerance parameter and the fraction of 
opposing neighbor is $f$.
The agents are satisfied if $u$ has a positive value or zero. 
Thus the utility factor not only carries the information whether the agent is
happy or not, it also tells {\it{how}} happy or unhappy they are. 
The moves are only
allowed for the unsatisfied agents and are  
 based on the value of $u$ in the old and new locations; movements are only made if $u$ increases in the new location.
Thus it is a constrained model.
 The discrete model is  retrieved  by  considering 
only the  sign of $u$ for making movements and the corresponding model is termed  model A. 
The truly continuous model where actual values of $u$ are taken to determine 
movements is called model B. Both models are studied to present a comparative picture. 
In model B, jumps are possible even if $u$ remains negative but is larger at the new site. 
This makes  agents in  model B  more mobile compared to model A and different from constrained models with discrete utility factors.

Model A and model B are shown to differ drastically as far as segregation, phase transition and other dynamical behaviour are considered. 
Model A is identical to the original Schelling model except for the fact that nonlocal jumps are allowed here. 
It is different from the model of \cite{gauvin} as only movement of unsatisfied
agents are considered although in both models nonlocal jumps are allowed. Model A shows a frozen state and a sharp transition 
to a segregated state at a nonzero value of $F$. The segregation factor $s$ 
remains close to $1$ before decreasing slowly to $0.5$, the value corresponding to a completely mixed state. Defining the segregated state 
to be that for which $s>0.9$, one can obtain a boundary between the segregated and the mixed state.

The most striking observation for model B is that here the frozen state does not exist in the thermodynamic limit in contrast to model A 
and unconstrained model with discrete utility \cite{gauvin}. 
One can justify the presence (absence) of the frozen state in model A (model B) for small $F$ in the following way. 
Let the two groups be labelled X and Y. Let at a particular vacant site all the neighboring sites be occupied (this will be more probable for small $p$) and $y$ be the number of Y type neighbors. 
Now the probability that an X type agent which is unsatisfied at its present position will jump to this particular vacant site is
$\sum_{y=0}^{8F}[{{8} \choose {y}}p(\frac{1-p}{2})^{8}]$ in model A. If $F$ is very small, then only very few terms will contribute to the sum.
Hence the probability is rather small and very few unsatisfied agents will be able to move in a finite time (note that only one attempt is allowed) for small $F$ and $p$. Thus for all practical purposes this makes the majority of the agents stay in their present unsatisfied state. 
In model B, the probability that an X type unsatisfied agent with $y'$ number of Y type neighbors in its present position 
will jump to this particular vacant site is
$\sum_{y'=(8F+1)}^{8}\sum_{y=0}^{y'-1}[{{8} \choose {y}}p(\frac{1-p}{2})^{8}]$.
Obviously this probability is much larger than that in model A and the frozen state ceases to occur in model B.

In model B, the segregation factor remains very close to $1$ from $F=0$ to 
a finite value of $F$ and then decreases slowly to $0.5$.
 One can obtain a boundary between a segregated state and mixed state as in the case of model A. 

 Since model A is identical to the  constrained Schelling model with 
discrete utility, it is not surprising 
that only small clusters are generated.
In model B, large clusters are formed similar to the unconstrained model. 
This is confirmed from the calculation of the correlation as a function of time. Of course, a rigorous calculation of cluster sizes as a function of the parameter $F$ and $p$ 
will be able to distinguish model A and model B more quantitatively and will be reputed in a future publication \cite{ppr}. 
Other features like persistence probability $p_{move}$ and average distance travelled $d$ also show sharp differences in the two models. In model A, these quantities are largely 
affected by the presence of the phase transition at $F_c$; there is  non-monotonic
behaviour in model A. 
We may add the remark here that although the behaviour of $p_{move}$ and $d$ have clearly different
behaviour in the two models, the variation with $F$ in model A for $F> F_{c}$ 
and in model B for $F > 0$ is quite similar for both quantities.  
Note that in these two regions $\langle u \rangle > 0$ and thus the agents are
happy on an average, which seems to be the relevant factor determining the trends of $p_{move}$ and $d$. 
It will be interesting to compare these findings with real data, if available, in future studies.

In a way, model B is a semi-constrained model as movements are less restricted here and perhaps it is not entirely surprising that the segregation occurs here to a larger
extent. However, the absence of the frozen state is a complete surprise.

The present model with continuous utility (model B) shows that moves which 
make utility larger, but not necessarily positive, helps in attaining segregated
states more effectively. In contrast to the unconstrained model, this is a more
realistic strategy as in the unconstrained model, the tendency to move keeping the 
utility factor unchanged does not seem to be practical. 
On the other hand, in the study of the constrained model B,
we get the important message that  even if one cannot achieve the ideal state
in a single step, it is worthwhile if an action brings one closer to it. 
 
Acknowledgement: The authors gratefully thank P. Gade for the extensive discussions which led to the formulation of the problem. 
They also acknowledge financial support from CSIR project (PS) and UGC fellowship (PR). Computations have been done 
in the HP cluster provided by DST (FIST) project.

\end{document}